\documentstyle{l-aa}       
\makeatletter
\def\cal{\fam\tw@}
\makeatother
\begin{document}
\thesaurus{12 (12.04.1; 12.07.1)}    
\title{Lens reconstruction and source redshift distribution}
\author{Pierre-Yves Longaretti}
\institute{
Laboratoire d'Astrophysique de l'Observatoire de Grenoble, B.P. 53,
F-38041 Grenoble Cedex 9, France (pyl@gag.observ-gr.fr)}
\offprints{P.Y. Longaretti (pyl@gag.observ-gr.fr)}
\date{Submitted to A\&A {\it Letters}}
\maketitle
\begin{abstract}
  The currently used linear and nonlinear lens inversion techniques are based
on distortion estimators whose complicated source redshift
dependence makes the influence of the redshift distribution of the sources
difficult to take into account and to analyze. However, the lens equations can 
be explicitly averaged over the source redshift distribution by a suitable 
choice of the estimators. Lens reconstruction procedures are outlined, 
which make use of all the information and in which all the unknown quantities 
of the problem can be recovered simultaneously in both the linear and nonlinear
regime, for either noncritical or critical lenses. These procedures require
no prior knowledge of the redshift distribution of the sources, but are 
possibly dependent on the Cosmology in some instances. Possible 
methods of recovery of the source redshift distribution itself are briefly
discussed at the same time.

\end{abstract}
\section{Introduction}
  A decade after the discovery of giant gravitational arcs in
distant clusters of galaxies (Soucail {\it et al.} 1987, Lynds and Petrosian
1986), the strong and weak lensing distortions induced by the distribution of
foreground dark matter on the position and shape of background sources have 
become one of the major tools
in modern observational Cosmology. Early work (e.g., Kochanek 1990, or 
Miralda-Escud\'e 1991, 1993) has focused on parametric reconstruction of the
lens properties; a significant progress along this line has been performed by
the identification and use of multiple images of a same source (e.g., Mellier 
{\it et al.} 1993, Kneib {\it et al.} 1993, 1995). Two major breakthroughs
have been made in the recent years: on the theoretical side, powerful linear
and nonlinear nonparametric lens reconstruction methods have been proposed
(e.g., Kaiser and Squires, 1993; Kaiser 1995; Seitz and Schneider 1995), while
on the observational side, new and impressive techniques have been developed
to measure the distortion field in the weak lensing regime from the distortion
of individual sources (Bonnet {\it et al.}  1994) and
from the autocorrelation function of whole images (Van Waerbeke {\it et al.}
1997). Finally the end of the long cosmological quest for the magic numbers
of the Universe ($\Omega$ and $\Lambda$; see, e.g., Fort {\it et al.} 1997) 
and even the determination of the power spectrum of large scale structures 
(Bernardeau {\it et al.} 1997) seem within reach of the lensing approach.

However, a remaining vexing difficulty plagues these attempts, namely the 
complicated dependence of the distortion estimators on the background source 
redshifts, which makes the influence of the source redshift distribution on
the lensing analyses difficult to quantify (see, e.g., Luppino and Kaiser, 
1996).  The purpose of this {\it Letter} is to
show how this difficulty is resolved by an appropriate choice of the distortion
estimators (this was already partially pointed out by Kochanek 1990; however, 
the implications for the elaboration of reconstruction procedures with a 
distribution of source redshifts seem to have gone unnoticed; for an 
alternative approach, see Seitz and Schneider 1997). 
Sect.\ 2 is devoted to an exposition of the lensing and deformation equations 
used in the paper, both to specify the adopted notation and introduce the 
distortion estimators. Sect.\ 3 begins with a
presentation of the relevant probability distributions of images and sources, 
and moves on to the statistical properties of the chosen distortion estimators.
Sect.\ 4 discusses these results and concludes this {\it Letter} by outlining 
possible methods of recovery of the lens projected mass distribution and of
the source redshift distribution for noncritical as well as critical lenses.
\section{Lensing equations and notations:}
   In what follows, we make use of quantities which can be defined in similar
ways for the sources, the images and the lens. Such quantities are accordingly
indexed with a subscript or a superscript ``$I$", ``$S$" and ``$L$". Latin 
subscripts refer to
the angular directions in the plane of the sky, which are labeled $x_1$ and
$x_2$. With these definitions, the lens equation reads
\begin{equation}
{\bf x}^{S}={\bf x}^{I}-{\bf\nabla}\phi({\bf x}^{I}), 
\end{equation}
\noindent where the bidimensional lensing potential $\phi$ is related to the
lens surface density $\Sigma$ by the bidimensional Poisson equation
\begin{equation}
\Delta\phi=2{\Sigma\over\Sigma_c}. 
\end{equation}
\noindent The critical surface density $\Sigma_c$ just introduced 
is defined by
\begin{equation}
\Sigma_c\equiv {c^2\over 2\pi G}{D^S\over D^{LS} D^L}, 
\end{equation}
\noindent where $D^{L}$, $D^{S}$, and $D^{LS}$ are the angular-diameter distance
of the observer to the lens, the observer to the source, and the lens to the
source, respectively.

  Extended objects are characterized by their surface brightness 
$I_S({\bf x}_S) = I_I({\bf x}^I)$, the position of their centers 
in the source plane and in the image plane
\begin{equation}
{\bf X}^{S,I}={\int\!\!\int\ I_{I,S}({\bf x}^{I,S}){\bf x}^{I,S}\ d^2 
{\bf x}^{I,S}\over
\int\!\!\int\ I_{I,S}({\bf x}^{I,S})\ d^2 {\bf x}^{I,S}}, 
\end{equation}
\noindent and their shape matrix
\begin{equation}
M^{I,S}_{ij}={\int\!\!\int\ I_{I,S}(x_i^{I,S}-X_i^{I,S})
(x_j^{I,S}-X_j^{I,S})\ d^2 {\bf x}^{I,S}\over
\int\!\!\int\ I_{I,S}\ d^2 {\bf x}^{I,S}}. 
\end{equation}
  Following Kneib {\it et al.} (1994), it is convenient to express the 
normalised source and image shape matrices $M/({\rm det} M)^{1/2}$ in 
terms of the orientation of the principal axes, $\theta$ (defined
such that $\tan (2\theta)=2M_{12}/(M_{11}-M_{22})$ ), and the major axis
$a_0$ and minor axis $b_0$ of the equivalent ellipse as
\begin{equation}
{M\over({\rm det} M)^{1/2}}=
         \left( \begin{array}{cc}
                \delta+|\tau|\cos (2\theta) & |\tau|\sin (2\theta) \\
                |\tau|\sin (2\theta) & \delta - |\tau|\cos (2\theta)
                \end{array} \right),                          
\end{equation}
\noindent where the modulus of the complex quantity 
$\tau\equiv|\tau|\exp 2{\rm i}\theta$ is given by
\begin{equation}
|\tau|  = {a_0^2-b_0^2\over 2a_0b_0},             
\end{equation}
\noindent and
\begin{equation}
\delta  =  {a_0^2+b_0^2\over 2a_0b_0} = (1+|\tau|^2)^{1/2}. 
\end{equation}
  Finally, one can define a last observable, $\sigma$, which measures the 
amplitude of the deformation matrix:
\begin{equation}
\sigma=8({\rm det} M)^{1/2}= 2ab.   
\end{equation}
  When the distortions are moderate (so that the amplification matrix can be
assumed to be nearly constant across the image), the source and image shape 
matrices $M$ are related by
\begin{equation}
M^S=a^{-1}M^I a^{-1}, 
\end{equation}
\noindent where the inverse of the amplification matrix $a^{-1}$ reads
\begin{equation}
a^{-1}= \left( \begin{array}{ll}
                1-\phi_{,11} & -\phi_{,12} \\
                -\phi_{,12} & 1-\phi_{,22}
                \end{array} \right).                          
\end{equation}
\noindent The coma designates partial derivatives with respect to coordinates
represented by their indices. 

It is worth pointing out here that Van Waerbeke {\it et al.} (1997) 
have introduced a different distortion estimator,
the second moments of the autocorrelation function in the image plane. 
This new matrix estimator possesses at least two 
fundamental advantages over the more conventional individual image shape
matrix: first, it makes maximal use of all the information contained
in CCD images, including the information contained in the noise,
so that this autocorrelation matrix can be extracted from the data everywhere
in the field with unprecedented accuracy, which makes it the most promising
data analysis tool to this date; furthermore, this estimator 
transforms like the shape distortion matrix
introduced above, so that the analysis developed below can be literally
transposed to it. 

  Introducing the convergence
\begin{equation}
\kappa={\Delta \phi\over 2}={\Sigma\over\Sigma_c}, 
\end{equation}
\noindent and the complex shear
\begin{equation}
\gamma \equiv |\gamma|\exp({\rm i}2\theta_L) =
              {1\over 2}(\phi_{,22}-\phi_{,11})-{\rm i}\phi_{,12}, 
\end{equation}
\noindent the inverse amplification matrix reads
\begin{equation}
a^{-1} = \left( \begin{array}{cc}
                1-\kappa+|\gamma|\cos (2\theta_L) & |\gamma|\sin (2\theta_L) \\
                |\gamma|\sin (2\theta_L) & 1-\kappa - |\gamma|\cos (2\theta_L)
                \end{array} \right).                         
\end{equation}
  Besides $\gamma$, two other quantities related to the lens potential, 
$\tau_L$ and $\delta_L$, are needed in the following analysis, and write
\begin{eqnarray}
\tau_L & = & {-2(1-\kappa)|\gamma|\over |A^{-1}|}=
      =  {-2(1-\kappa)|\gamma|\over |(1-\kappa)^2-|\gamma|^2|},\\ 
\delta_L & = & {(1-\kappa)^2+|\gamma|^2\over |A^{-1}|}=
      =  {(1-\kappa)^2+|\gamma|^2\over |(1-\kappa)^2-|\gamma|^2|}, 
\end{eqnarray}
\noindent where $A^{-1}={\rm det}(a^{-1})$. Note that $\delta_L^2=1+\tau_L^2$, 
as the equivalent source and image quantities.

  With these definitions, the shape matrix transformation equation [Eq. (10)]
yields
\begin{eqnarray}
\sigma_I & = & {\sigma_S\over |A^{-1}|},\\  
|\tau_S|\sin(2\theta_S-2\theta_L) & = & |\tau_I|\sin(2\theta_I-2\theta_L),\\
|\tau_S|\cos(2\theta_S-2\theta_L) & = & \delta_L|\tau_I|
\cos(2\theta_I-2\theta_L)-\tau_L\delta_I.    
\end{eqnarray}
\noindent  The last equation can be inverted, which gives 
\begin{equation}
|\tau_I|\cos(2\theta_I-2\theta_L) =  \delta_L|\tau_S|\cos(2\theta_S-2\theta_L)+
\tau_L\delta_S,    
\end{equation}
  In the rest of the paper, the complex distortion parameter $\tau$ and
the distortion amplitude $\sigma$ are used instead of the shape matrix.

  These distortion equations present several advantages over the ones which
are presently used in inversion procedures. First, the linear relation between
source and image distortion estimators simplifies greatly the averaging over
the source orientation. Second, the $|A^{-1}|$ denominator makes 
an explicit redshift averaging of the equations possible, as already
pointed out by Kochanek (1990), and shown in Sect. 3. On the other hand, 
$\tau$ is sensitive to the intrinsic ellipticity distribution of the sources, 
so that a single highly elongated source can in principle bias the estimation
of the local averages used in the next sections; however, this seems likely
to happen only rarely in practice for reasonable source redshift distributions.
\section{Probability distributions and distortion statistics:}
  The lens and the distortion equations involve both deterministic paramaters
(the lens redshift and mass distribution, the cosmological parameters) and
random variables (the position ${\bf x}^S$, the complex distortion ${\bf 
\tau}_S$, the distortion amplitude $\sigma_S$ and the redshift $z_S$ for the 
sources, the position ${\bf x}^I$, the complex distortion ${\bf \tau}_I$ and
the distortion amplitude $\sigma_I$ for the images; the number of sources and
images per unit area is also a random variable, but it is not used here). 
It is commonly assumed 
that the positions, distortion estimators, and redshifts of the sources (i.e., 
in the absence of the lens) are all independent random variables (however, the
following argument is unaffected by a possible correlation between $\sigma_S$
and $|\tau_S|$). Furthermore,
the positions and orientations $\theta_S$ of the sources are assumed to be 
uniformly distributed random variables. 

  With the assumption just recalled, the joint probability density distribution
of source parameters $p({\bf x}^S, \tau_S, \sigma_S, z_S)=
(1/4\pi)p(\tau_S)p(\sigma_S)p(z_S)$. The
statistical quantities of interest are the averages of the distortion
estimators over the source random variables, at a given position
in the lens plane, i.e., the probability distribution needed is the
conditional probability $p(\tau_S, \sigma_S, z_S/{\bf x}^I)\equiv p({\bf x}^I,
\tau_S, \sigma_S, z_S)/p({\bf x}^I)=
p({\bf x}^I, z_S)p(\tau_S)p(\sigma_S)/p({\bf x}^I)$.
Furthermore, 
\begin{equation}
p({\bf x}^I, z_S)={1\over 4\pi}p(z_S)|A^{-1}|,         
\end{equation}
\noindent from which one obtains
\begin{equation}
p({\bf x}^I)={1\over 4\pi}\langle|A^{-1}|\rangle_{z_S},     
\end{equation}
\noindent where $\langle |A^{-1}|\rangle_{z_S}$ stands for the redshift average
of $|A^{-1}|$ at given image position ${\bf x}^I$. Finally, the required 
probability distribution reads
\begin{equation}
p(\tau_S, \sigma_S, z_S/{\bf x}^I)=p(\tau_S)p(\sigma_S)p(z_S){|A^{-1}|\over
\langle |A^{-1}|\rangle_{z_S}}     
\end{equation} 
  For critical lenses, $\int\ p({\bf x}^I)\ d^2{\bf x}^I > 1$, because 
multiple images are absolutely correlated events and should not be coadded 
as independent ones. If multiplied by the number of sources in the source 
plane, Eq. (22) yields the expectation value of the number density in the 
image plane. 

Averaging the distortion equations for $\delta_I$, the complex distortion 
$\tau_I$ and the distortion amplitude $\sigma_I$ over the sources at a given 
position in the image plane now yields
\begin{eqnarray}
\langle\sigma_I\rangle & = & {\langle\sigma_S\rangle\over
\langle|A^{-1}|\rangle_{z_S}},\\   
\langle\delta_I\rangle
& = & \langle \delta_S\rangle_{|\tau_S|}{\langle\delta_L |A^{-1}|\rangle_{z_S}
\over\langle|A^{-1}|\rangle_{z_S}},
\\     
\langle\tau_I\rangle 
 & = & \langle \delta_S\rangle_{|\tau_S|} 
{\langle\tau_L |A^{-1}|\rangle_{z_S} 
\over\langle|A^{-1}|\rangle_{z_S}}
{\gamma\over|\gamma|},   
\end{eqnarray}
\noindent where $\tau_I=|\tau_I|\exp(2{\rm i}\theta_I)$, and where
the ratio $\gamma/|\gamma|$ does not depend on the redshifts of the sources. 
The left-hand sides of Eqs. (24) through (26) are provided by the observations.

  It is necessary to express the redshift averages appearing in the preceding
equations in more detail. To this purpose, let us introduce the lens critical
radius $b\equiv D_{LS}/D_S$, and define a redshift independent convergence
$K$ and shear $\Gamma$ by 
\begin{equation}
\kappa\equiv b K,   
\end{equation}
\begin{equation}
\gamma\equiv b \Gamma,   
\end{equation}
\noindent so that
\begin{equation}
\langle\tau_L |A^{-1}|\rangle_{z_S}=-2\langle b\rangle|\Gamma| + 
2\langle b^2\rangle K|\Gamma|,   
\end{equation}
\begin{equation}
\langle\delta_L |A^{-1}|\rangle_{z_S}=1-2\langle b\rangle K +\langle b^2\rangle
(K^2+|\Gamma|^2).    
\end{equation}
\noindent The sign of $A^{-1}$ depends on the position in the image plane and 
on the source redshift. If the sources are all at the same redshift, the
sign change occurs on the critical lines, i.e. a zero-width region. If the
sources have some dispersion in redshift, the critical lines become critical
zones. Two cases naturally emerge, depending on the size of these critical 
zones.

   If the critical zones are too small to be of practical importance, one has
\begin{equation}
\langle |A^{-1}|\rangle_{z_S}={\rm sgn}(A^{-1})\left[1-2\langle b\rangle K + 
\langle b^2\rangle (K^2-|\Gamma|^2)\right],  
\end{equation}
\noindent where ${\rm sign}(A^{-1})$ is well-defined outside the (neglected)
critical zones.
Both $\langle b\rangle$ and $\langle b^2\rangle$ 
depend only on the distribution of source reshifts and on the Cosmology.
Also,  
$\langle\delta_I\rangle^2/\langle\delta_S\rangle^2 = 1 + 
|\langle\tau_I\rangle|^2/\langle \delta_S\rangle^2$.

Otherwise, at
every position ${\bf x}^I$ one can define a redshift domain $Z^+$ for which
$A^{-1} > 0$, and a complementary redshift domain  $Z^-$ where $A^{-1} < 0$;
these domains are known if the lens surface density is known, for a given
Cosmology. Let us define $B_n^{\pm}=\int_{Z^{\pm}}\ b^n p(z_S)\ d z_S$; one has 
$B_n^+ + B_n^-=\langle b^n\rangle$ ($=1$ for $n=0$), where $\langle b\rangle$
and $\langle b^2\rangle$ have the same (position independent) value as in 
regions of constant sign of $A^{-1}$ (e.g., outside the asymptotic external
critical line). Then
\begin{equation}
\langle |A^{-1}| \rangle_{z_S} = (\Delta B_0) - 2(\Delta B_1)K + 
(\Delta B_2)(K^2-|\Gamma|^2),    
\end{equation}
\noindent where $\Delta B_n = B_n^+ - B_n^-$. Note that the three quantities 
$\Delta B_n$ are all related to each other through the source redshift 
distribution and the Cosmology, so that the spatial dependence of 
$\langle |A^{-1}|\rangle_{z_S}$ provides us with an integral 
equation for $p(z_S)$, for a given Cosmology and for a given lens surface 
density distribution. As $Z^{\pm}$ cover the whole redshift range of the 
distribution when one crosses the critical zones, this integral can in 
principle be solved for $p(z_S)$, possibly in a parametric way.
\section{Reconstruction procedure and discussion:}
  Following a lead by Kaiser and Squires (1993), Seitz and Schneider (1995)
have shown that the convergence and the complex shear are related through an
invertible convolution equation which yields
\begin{eqnarray}
\Gamma({\bf x}^I) & = & {1\over\pi}\int d^2{\bf y}^I\ G({\bf x}^I-{\bf y}^I) 
K({\bf y}^I),\\  
K({\bf x}^I) & = & {1\over\pi}\int d^2{\bf y}^I\ {\rm Re}\left(G({\bf x}^I-
{\bf y}^I) \Gamma^*({\bf y}^I)\right),  
\end{eqnarray}
\noindent where the kernel $G$ is given by
\begin{equation}
G({\bf x})={x_1^2-x_2^2+2{\rm i}x_1 x_2\over |{\bf x}|^4}  
\end{equation}
 Two different cases must be distinguished, depending on whether one can ignore
the critical zones defined above. Let us first assume that
we can, which is probably correct in a majority of cases.
Then Equations (24) through (35) 
[especially Eqs. (24), (26) and (33)] provide us with enough independent 
constraints for the five quantities ($K$, ${\rm Re}(\Gamma)$, 
${\rm Im}(\Gamma)$, $\langle b\rangle$ and 
$\langle b^2\rangle$), which can therefore in principle be solved for as 
functions of ${\bf x}^I$ (with appropriate noise filtering; Kaiser and Squires,
1993, and Seitz and Schneider, 1995). Note that 
$\langle |A^{-1}|\rangle_{z_S}$ is obtained from the
distortion amplitude $\sigma$ but any other estimator of $|A^{-1}|$ would
do as well (e.g., it is possible that the magnification might be extracted with
more accuracy from the image autocorrelation itself than from its second 
order moments). Also, $\sigma$ and $\delta$ are expected to be
observationally less well constrained than $\tau$. However, the expected 
constancy of $\langle b\rangle$ and $\langle b^2\rangle$ in the image plane
provides us with an additional constraint on the inversion procedure,
as will the identification of multiple images, if any, so that,
although this inversion procedure would probably be difficult to implement 
from the observed distortions of individually identified objects alone, it 
seems within reach of the autocorrelation method proposed by Van Waerbeke {\it 
et al.} (1997). Note however that the strong amplification near the critical
zones biases the observed distribution of objects towards fainter or farther
sources, which can possibly invalidate the assumed constancy of $\langle b
\rangle$ and $\langle b^2\rangle$.

Note also that the local degeneracy pointed out by Schneider and Seitz (1995) 
is broken as long as $\langle b^2\rangle$ is significantly different from
$\langle b\rangle^2$. As expected, the global degeneracy of the distortion
equations
cannot be raised without the use of external information. In the same vein,
one also needs an external constraint to determine the magnitude of the
potential, as Eqs. (24) through (35) are invariant under the scaling
$K, \Gamma \rightarrow \alpha K, \alpha \Gamma$, and
$\langle b\rangle,\langle b^2
\rangle \rightarrow \langle b\rangle/\alpha, \langle b^2\rangle/\alpha^2$
(within the limits of variation of $b$). 
The simplest such constraint is obtained from the knowledge of the redshift
of a giant arc (if one is present in the field, which fortunately is not an 
exceptional situation). As for the source redshift distribution, 
$\langle b\rangle$ and $\langle b^2\rangle$ depend
on both the redshift of the lens and on the Cosmology besides the shape of the
source redshift distribution, so that it seems likely that the combination of 
the application of the reconstruction procedure just outlined to a number of
lenses at different (known) redshifts with other 
lensing methods depending on the characteristics of the redshift distribution
and constraining the Cosmology (see, e.g., Fort {\it et al.} 1997, and 
Bernardeau {\it et al.} 1997) will finally yield both the source redshift 
distribution and the cosmological parameters $\Omega$ and $\Lambda$. 

  Although it seems less likely to occur in practice, let us now also
consider the other possibility, namely that the critical zones are
extended enough in the image plane so that they
cannot be ignored in the reconstruction procedure. Comments made above and
applying to this case (with appropriate transpositions) are not 
repeated here. It is likely that the 
inversion is still possible in principle, although the reconstructed surface
density becomes Cosmology dependent. For example, one can assume an a priori
Cosmology and redshift distribution, which fixes the $\Delta B_n$ factors in
Eq. (32). Then Eq. (32) depends only on the lens mass distribution, which can 
be recovered in much the same way as described above along with $\langle b
\rangle$ and $\langle b^2\rangle$ (if needed) which can be considered as free 
parameters for the mass distribution reconstruction. Then Eq. (32) can be 
solved for the redshift distribution for a given Cosmology and the previously
found surface density and possibly the values of $\langle b\rangle$ and 
$\langle b^2\rangle$, and the procedure 
iterated until convergence is obtained. The weak points of this approach are 
that the limited size of the concerned regions in the image plane, the
possible breakdown of the approximation underlying Eq. (10) for some sources
not individually identified, the nonvanishing size of the background sources, 
the limited background sources number density and other observational sources 
of noise and bias are likely 
to make such a reconstruction procedure much more difficult to implement 
than the preceding
one (not mentioning possible problems in the reconstruction of $p(z_S)$ from
the integral equation). The strong point is that a single well-chosen lens, 
combined with the methods mentioned above to constrain the Cosmology could in
principle yield at the same time the cosmological parameters and the source
redshift distribution, a rather tantalizing possibility.
\begin{acknowledgements}
  I thank Yannick Mellier and Peter Schneider for informal but informative
discussions on the content of the paper.
\end{acknowledgements}

\end{document}